\documentclass[
    superscriptaddress,
    showpacs,
    preprintnumbers,
    bibnotes,
    amsmath,
    twocolumn,
    amssymb,
    aps,
    prx,
]{revtex4-1}

\usepackage{setspace}
\usepackage{enumitem}
\usepackage{graphicx}    
\usepackage{dcolumn}    
\usepackage{bm}             
\usepackage{color,soul}          
\usepackage{amssymb,amsfonts,amsmath,amstext}

\newcommand{\ba}{\begin{eqnarray}}
\newcommand{\ea}{\end{eqnarray}}
\renewcommand{\v}[1]{{\boldsymbol #1}}

\begin{document}

\title{Type-II Dirac line node in strained Na$_3$N}

\author{Dongwook Kim}
\affiliation{Department of Physics, Sungkyunkwan University, Suwon 16419, Korea}
\author{Seongjin Ahn}
\affiliation{Department of Physics and Astronomy, Seoul National University, Seoul 08826, South Korea}
\author{Jong Hyun Jung}
\affiliation{Department of Physics and Astronomy, Seoul National University, Seoul 08826, South Korea}
\author{\\ Hongki Min}
\affiliation{Department of Physics and Astronomy, Seoul National University, Seoul 08826, South Korea}
\author{Jisoon Ihm}
\affiliation{Department of Physics, Pohang University of Science and Technology, Pohang 37673, South Korea}
\author{Jung Hoon Han}
\affiliation{Department of Physics, Sungkyunkwan University, Suwon 16419, Korea}
\author{Youngkuk Kim}
\email{youngkuk@skku.edu}
\affiliation{Department of Physics, Sungkyunkwan University, Suwon 16419, Korea}
\affiliation{Center for Integrated Nanostructure Physics, Institute for Basic Science (IBS), Suwon 16419, Korea}
\date{\today}

\begin{abstract}
Dirac line node (DLN) semimetals are a class of topological semimetals that
   feature band-crossing lines in momentum space.  We study the type-I and
   type-II classification of DLN semimetals by developing a criterion that
   determines the type using band velocities.  Using first-principles
   calculations, we also predict that Na$_3$N under an epitaxial tensile strain
   realizes a type-II DLN semimetal with vanishing spin-orbit coupling (SOC),
   characterized by the Berry phase that is $\mathbb{Z}_2$-quantized in the
   presence of inversion and time-reversal symmetries.  The surface energy
   spectrum is calculated to demonstrate the topological phase, and the type-II
   nature is demonstrated by calculating the band velocities. We also develop a
   tight-binding model and a low-energy effective Hamiltonian that describe the
   low-energy electronic structure of strained Na$_3$N.  The occurrence of a
   DLN in Na$_3$N under strain is captured in the optical conductivity, which
   we propose as a means to experimentally confirm the type-II class of the DLN
   semimetal.  
\end{abstract}

\maketitle

\section{Introduction}
During the past decade, topological semimetals have been an active subject of
condensed matter physics and quantum materials, providing a unique venue for
exotic phenomena originated from band topology \cite{Arnold2016, Zhang2016,
Li2016, He2014}. Since the discovery of Weyl semimetals \cite{Burkov2011,
Wan2011, Yang2011, Xu2011, Singh2012, Huang2015, Xu613, Xu294, Weng2015,
Lv2015, Xu2016, Xu2015, Jia2016}, various classes of topological semimetals
have been found theoretically and experimentally, including Dirac semimetals
\cite{Xu560,Liu864,Wang2012,Young2012,Borisenko2014, Wang2013, Liu2014,
Kim15p036806, Yu15p036807}, nodal line semimetals \cite{Burkov2011,
yu2016topological, fang2016topological, weng2016topological, Chiu16p035005,
Fang15p081201, Kim15p036806, Yu15p036807, Xu17p045136,
Huang16p201114,Young17p085116, Zhao16p195104, Du17p3, Weng15p045108,
Chen15p6974, Wang16p195501, Cheng16p468, Ezawa16p127202,
Hirayama17p14022,Li16p096401, Gan16p161106386, Weng16p165201,Zhu16p031003,
Lu16p160304596, Xie15p083602}, and their diverse variations such as multi-Weyl
semimetals and double Dirac semimetals \cite{Fang2012, Wieder2016, Huang1180,
Bradlynaaf5037, Chang2017}.

More recently, the classification of topological semimetals is further
specified into the type-I and type-II classes based on the geometric structure
of the Fermi surface \cite{Soluyanov2015, Wang2016, Deng2016,Yazyev2016,
Kim2017, Yan2017, Huang2016, Chang2017}. The type-I material features a closed
Fermi surface enclosing the nodes; in contrast, type-II material features open
Fermi surface composed of electron and hole pockets that linearly touch at the
nodes. First proposed in Weyl semimetals, the type-I/II classification has been
extended to Dirac semimetals \cite{Zhang2017, Noh2017, Fei2017, Kim2017,
Guo2017, Chang2017} and nodal line semimetals \cite{Li2017, 1711.09167,
1709.08287}. In the case of Weyl semimetals, such classification provides an
important insight for understanding unique phenomena present only in the
type-II materials, such as the squeezing or collapse of the Landau levels, the
Klein tunneling, and magnetic breakdown in momentum space when magnetic fields
are applied to over-tilted Weyl nodes \cite{O'Brien2016, Yu2016p,
Tchoumakov2016, 1708.09387}.  Some papers have already proposed type-II DLNs
and reported relevant materials hosting such kinds of DLNs \cite{Li2017,
1711.09167, 1709.08287}.  Contrasting features of type-I and type-II DLNs have
been identified in terms of their dispersion and Fermi surface geometry
\cite{Li2017}.

In this paper we go beyond the well-known characterization of type-II DLNs and
identify a more fundamental way to understand and identify the type-II class of
DLN semimetals. Based on geometric argument, a concrete connection is
established between the sign inversion of band velocities around a type-II DLN
and the occurrence of open Fermi surface in type-II DLN semimetals, which leads
to a rigorous criterion to determine the types of DLNs.  We use this criterion
and first-principles calculations to predict that epitaxially strained Na$_3$N
realizes a type-II DLN semimetal phase.  The nontrivial band topology and
type-II nature of the DLN semimetal are explored. We also construct a
tight-binding model and a low-energy effective theory, and use them to
rationalize the low-energy electronic structure and to calculate the optical
conductivity of Na$_3$N.  The optical response significantly changes upon
straining due to the occurrence of a type-II DLN semimetal, which can be
experimental evidence of the type-II DLN semimetal phase hosted in strained
Na$_3$N.

This paper is structured as follows.  First, we begin with the background in
Sec.\,\ref{sec:backgrounds}, which provides simple and intuitive ways to
understand the type-I versus type-II nature of DLNs.  Following that, we
contrast in Sec.\,\ref{sec:methods} the electronic band structures of pristine
and strained Na$_3$N. This clarifies the electronic structure of Na$_3$N that
is responsible for the occurrence of a DLN under strain.  Next, in
Sec.\,\ref{sec:DLN}, topological characterization of strained Na$_3$N is
discussed in terms of $\mathbb{Z}_2$ topological indices and surface energy
spectrum.  Also, the type-II nature is demonstrated via the band velocities
evaluated in the vicinity of the DLN.  Then, in Sec.\,\ref{sec:tb}, we
construct a tight-binding model for Na$_3$N, reproducing main features of the
DFT calculations.  Physical manifestations of the type-II DLN nature are
predicted to occur as unique features in the surface spectrum and the optical
conductivity, as demonstrated in Sec.\,\ref{sec:physical}.  Finally, we
conclude the paper with a summary and perspective in Sec.\,\ref{sec:summary}.

\begin{figure*}[!ht]
\centering
\includegraphics[width=\textwidth]{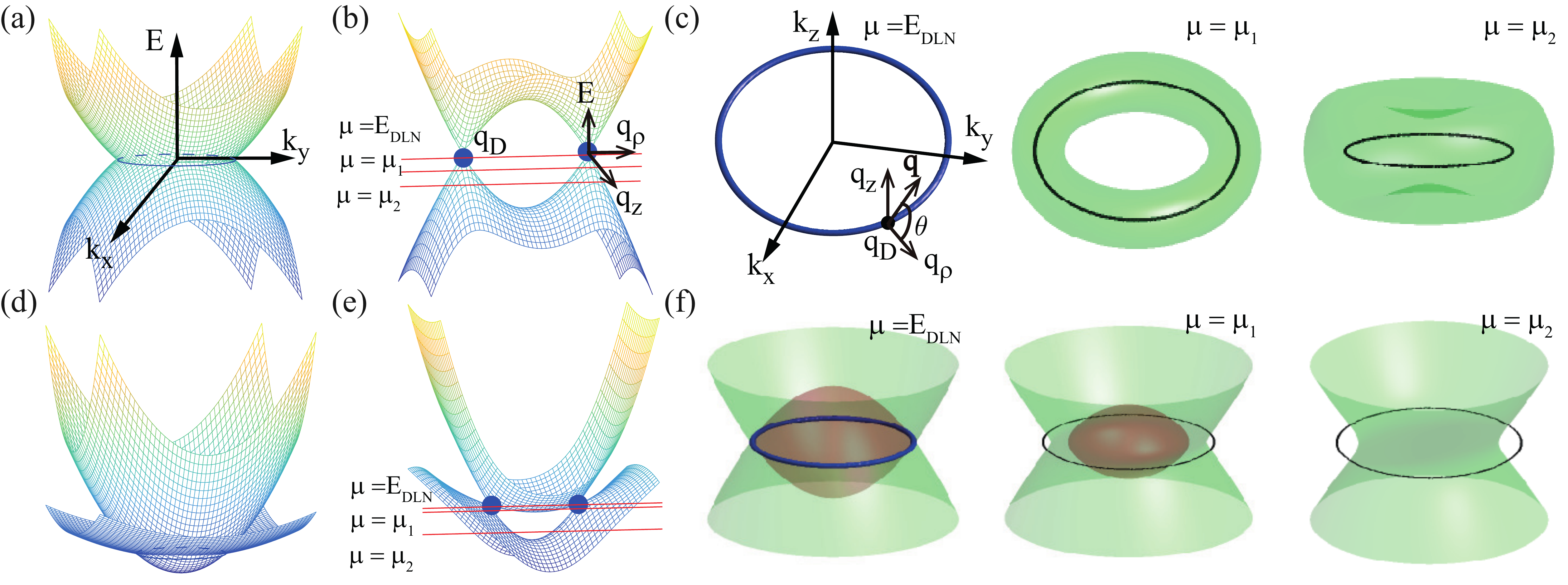} \\
\caption{The typical energy band dispersions and the Fermi surfaces of type-I
   (first row) and type-II (second row) DLNs. (a) and (b) illustrate the
   electronic energy bands of a type-I DLN on the $k_x - k_y$ and $k_\rho -
   k_z$ planes, respectively. Here  $k_\rho = \sqrt{k_x^2 + k_y^2}$ and the DLN
   lies on the $k_x - k_y$ plane. (c) Changes of the Fermi surface for a type-I
   DLN in 3D momentum space as chemical potential is tuned from $\mu = E_{\rm
   DLN}$, $\mu_1$ to $\mu_2$. The locations of $E_{\rm DLN}$, $\mu_1$, and
   $\mu_2$ in energy are shown in (b). The blue circles in (c) indicate the
   position of the DLN. The green surfaces represent the Fermi surface at the
   corresponding chemical potentials. $\v {q} = \v {k} - \v {k}_\mathrm{D}$ is
   the relative momentum defined with respect to a point of the DLN at $\v
   {k}_\mathrm{D}$.  The left panel of (c) defines the in-plane $q_\rho$ and
   out-of-plane $q_{z}$ components of the relative momentum $\v q$. (d) and (e)
   are the typical energy bands of a type-II DLN on the $k_x - k_y$ and $k_\rho
   - k_z$ planes. (f) illustrates the corresponding changes of the Fermi
   surface for a type-II DLN at $\mu = E_{\rm DLN}$, $\mu_1$, and $\mu_2$. The
   blue circles in (f) indicate the position of the DLN. The green (red)
   surfaces represent the electron (hole) pockets. At $\mu = E_\mathrm{DLN}$,
   the DLN coexists with the electron and hole pockets that touch each other
   linearly at the DLN.} \label{fig:fig1}
\end{figure*}

\section{Backgrounds}
\label{sec:backgrounds}

We start with a brief review of the mathematical foundation for the type-II DLN
semimetals. Similar to the Weyl semimetal case \cite{Soluyanov2015},
distinguishing aspects of type-II DLNs lie in their dispersion and the Fermi
surface geometry: the existence of over-tilted Dirac cone and the open (or
extended) Fermi surface \cite{Li2017}. Figure\,\ref{fig:fig1}(a) and (d)
illustrate the contrasting shape of energy bands of type-I and type-II DLNs on
the $k_x-k_y$ plane where the line node lies. Also, Fig.\,\ref{fig:fig1}(b) and
(e) show contrasting energy bands of type-I and type-II DLNs featured on the
$k_{\rho}-k_z$ plane for a fixed value of in-plane angle $\tan^{-1}(k_y /k_x)$,
where $k_\rho =\sqrt{k_x^2 + k_y^2}$ is the radial momentum. The $k_{\rho}-k_z$
plane intersects with the DLN at two points of the DLN. The intersecting points
of the DLN and the $k_{\rho}-k_z$ plane are indicated by blue dots in
Figs.\,\ref{fig:fig1}(b) and (e). These intersecting points appear as Dirac
points with linear dispersions away from the points on this normal plane.
Hereafter, we refer to them as Dirac points.

While the energy dispersions are linear in both type-I and type-II DLNs in the
vicinity of the nodal line, the difference between them is clear in the band
velocities at the DLN. Recall that two independent directions exist in moving
away from a given point on the DLN, which define a plane locally normal to the
tangent direction of the DLN. If the band velocities are opposite for the
conduction and the valence bands regardless of the direction away from the
Dirac point in this normal plane, we refer to this as a \textit{conventional}
type-I DLN. In contrast, if the band velocities of the conduction and valence
band have the same sign for some direction away from the DLN in the normal
plane, we refer to it as an \textit{unconventional} type-II DLN. These
different types give rise to the second distinguishing characteristic,
manifested in the Fermi surface geometry. As illustrated in
Fig.\,\ref{fig:fig1}(c) and (f), as one varies the chemical potential $\mu$
[red lines in Fig.\,\ref{fig:fig1}(b) and (e)], hole (electron) pocket appears
as a green (red) surface. Irrespective of the chemical potential $\mu$, the
Fermi surface (equally, constant energy surface at $E=\mu$) of a type-I
(type-II) DLN exists in a closed (open) surface, in line with the previous
study in \cite{Li2017}.

Instead of inspecting the band velocities, the classification is also possible
by the Fermi surface geometry, similar to the case of Weyl points
\cite{Soluyanov2015}. To specify this, we first introduce two coordinate
systems used in the discussion throughout the paper.  The first one is the $\v
k = (k_x , k_y, k_z)$ coordinates in the Brillouin zone (BZ) describing the
global structure of a DLN. The second set of coordinates is a relative
coordinate system $\v q = ( q_x, q_y, q_z)$ with the origin at a Dirac point of
the DLN $\v k_\textrm{D}$ [Fig.\,\ref{fig:fig1}(c)], which describes the local
behavior of the energy bands. The two coordinate systems are related by $\v k =
\v k_D + \v q$, where $\v k_\textrm{D}$ refers to a point on the DLN.

In general, the nodal structure of two bands can be described by $2 \times 2$ Hamiltonian
\begin{eqnarray}
   H = \sum_{j=1}^{3}d(\v q)_j \sigma_j,
\label{eq:eq0}
\end{eqnarray}
with the Pauli matrices $\sigma_j$'s.  When the Hamiltonian is linear in $\v
q$, one can find a linear map $\v A: V \rightarrow  W$ from $\v q$ to $\v d$,
where $ \v q \in V$, $\v d \in W$, and

\begin{eqnarray}
H = \sum_{j=1}^{3}d_j \sigma_j= \sum_{i,j=1}^{3}q_iA_{ij} \sigma_j .
\label{eq:eq1}
\end{eqnarray}
Band gap clossing points generically occur at the $\v q$-points that satisfy
$d_j(\v q) = 0$ for all $j$'s. Depending on the dimensions of the null space
(or kernel space) of the linear map $\v {A}$, which can be either 0, 1, or 2,
the gap closing points constitute points, lines, or  planes, respectively. The
kernel dimensions determine the dimensions of the quotient space
$V/\text{ker\textbf{A}}$ and also the dimensions of the image space of the
linear mapping $\text{im\textbf{A}}$ as $\text{Dim}(V) -
\text{Dim(ker\textbf{A})}$. We can rewrite the low-energy Hamiltonian using the
new linear mapping $\mathbf{\v {\tilde{A}}}$ from $V/\text{ker\textbf{A}}$ to
$\text{im\textbf{A}}$ with reduced dimensions ($\mathbf{\v {\tilde{A}}} :
V/\text{ker\textbf{A}} \rightarrow  \text{im\textbf{A}}$, $\v {k} \in
V/\text{ker\textbf{A}}$ and $\v {d} \in \text{im\textbf{A}}$). Now this new
linear mapping describes only the linear dispersion in the two directions that
are locally orthogonal to the nodal line direction.  As a result, in the case
of nodal line [Dim $\text{ker\textbf{A}}$=1], the local behavior of electrons
is described by the low-energy Hamiltonian with the reduced dimensions. [Dim
$V/\text{ker\textbf{A}}$=2]

\begin{eqnarray}
H = \sum_{i,j=1}^{2} q_i \tilde{A}_{ij} \sigma_j + \sum_{i=1}^{2} w_i q_i .
\label{eq:eq2}
\end{eqnarray}
Here ``local" means that the wave vector $\v q$ is expanded from a particular
point on the nodal line [See Fig.\,\ref{fig:fig1}(c)]. Representing the
position of a point constituting the DLN as $\v k_D$, wave vectors in the BZ
are expanded up to the linear order of $\v q = \v k - \v k_D$, leading to the
low-energy Hamiltonian Eq.\,(\ref{eq:eq2}) that describes a Dirac cone in two
dimensions. The second term in the Hamiltonian is introduced to describe the
tilting of the Dirac cones, which plays a crucial role enabling the type-II
DLNs. As in the case of the Weyl point (WP) \cite{Soluyanov2015}, the Fermi
surface equation, which is a conic section equation in the DLN case, gives
either an open or a closed solution, which defines the type-I/type-II
classifications.

In detail, two energy eigenvalues are obtained from Eq.\,(\ref{eq:eq2}) as
\begin{eqnarray}
E_\pm = \sum_{i=1}^{2}w_iq_i \pm \sqrt{\sum_{i,j=1}^{2}q_i[\tilde{A}\tilde{A}^T]_{ij}q_j} .
\label{eq:eq3}
\end{eqnarray}
The Fermi surface equation that satisfy $E_\pm = \mu$ is obtained as
\begin{eqnarray}
\sum_{i=1}^{2}q_i(w_iw_j - [\tilde{A}\tilde{A}^T]_{ij})q_j - 2\mu\sum_{i=1}^{2}w_iq_i + \mu^2 = 0 .
\label{eq:eq4}
\end{eqnarray}
When det($\v {w} \otimes  \v {w} - [\tilde{A}\tilde{A}^T]$) $> (<)$ 0 , Eq.
(\ref{eq:eq4}) describes a conic section equation that gives an elliptic (a
hyperbolic) curve. Thus, DLNs are classified into two types: type-I for the
elliptic (closed) curve and type-II for parabolic (open) curve. It should be
worthwhile to mention that, in general, the tilting coefficient $\v w$ can be a
function of $\v k$, although it is treated as a constant for simplicity in the
above derivation. When the tilting coefficient varies, such that det($\v {w}
\otimes  \v {w} - [\tilde{A}\tilde{A}^T]$) changes the sign along the DLN,
which can happen in highly anisotropic systems that reside near the transition
between this type-I and type-II, the sign change of the determinant identifies
a novel-type hybrid DLN \cite{PhysRevB.97.155152}. While this case shares the
characteristic features of type-II DLNs hosting the electron and hole pockets
that coexist with the DLN, it also allows for diverse possibilities in the
Fermi surface geometry, which could be of a particular interest with respect to
quantum oscillations and transport phenomena.

Another way of sorting out the types of DLNs is by looking at the sign
inversion of band velocities. Since the matrix $\v {\tilde{A} \tilde{A}^T}$ is
Hermitian with non-negative eigenvalues $\lambda_1^2 , \lambda_2^2 > 0$ (we
assume the generic situation of non-zero eigenvalues), the energy dispersion
after the rotation to the principal axis becomes

\begin{eqnarray}
E_\pm = w_1 q_1 + w_2 q_2 \pm \sqrt{ ( \lambda_1 q_1 )^2 + ( \lambda_2 q_2 )^2 } .
\label{eq:eq5}
\end{eqnarray}
After the rescaling of variables $q_1 \rightarrow q_1 /\lambda_1$, $q_2
\rightarrow q_2 /\lambda_2$ and introducing the polar coordinates $\v q = (q_1
, q_2 ) = q (\cos \theta_q, \sin \theta_q )$ one can further simplify the
energy equation [Eq.\,(\ref{eq:eq5}) ] to \ba 
E_\pm & = &  \pm v_{\pm} q  , \nonumber \\
v_\pm & = & 1 \pm \sqrt{{\frac{w_1^2}{\lambda_1^2} } + { \frac{w_2^2}{\lambda_2^2}}} \cos [ \theta_{\v q} ]. 
\label{eq:Epm1}
\ea
Here, $\theta_{\v q}$ measures the angle from one of the principal axes. The
conventional type-I Dirac cone shown in Fig.\,\ref{fig:fig1}(b) is realized if
the radial velocity satisfies $v_+ v_-  > 0$ for all orientations $\theta_{\v
q}$, or equivalently, when
\ba 
1 - \sqrt{{\frac{w_1^2}{\lambda_1^2}} + { \frac{w_2^2}{\lambda_2^2} } } > 0 . 
\label{eq:Epm}
\ea
Otherwise, the unconventional type-II Dirac cone shown in
Fig.\,\ref{fig:fig1}(e) is realized with $v_+ v_- < 0$ over some range of
$\theta_{\v q}$. After some unraveling of the algebra, one can prove that
Eq.\,(\ref{eq:Epm}) is equivalent to the condition ${\rm det} (\v {w} \otimes
\v {w} - [\tilde{A}\tilde{A}^T]) > 0$. This proves that the classification
scheme in terms of the conic section is equivalent to the one in terms of the
sign inversion of the band velocity.

The difference in the type-I and type-II band structures can be understood in
the language of differential geometry as well. The band velocities along the
principal axes are the well-known principal curvatures in differential
geometry, and their product is the Gaussian curvature that indicates the shape
of the surface.  For either the conduction band or the valence band [$E_\pm$ in
Eq.\,(\ref{eq:Epm1})], coordinates along the principal axes are $(q_1, q_2)$ and
the corresponding velocities (or principal curvatures) are given by $v_{\pm,
1(2)} = \partial E_\pm / \partial q_{1(2)}$, respectively. The Gaussian
curvatures for each band follows from $K_\pm = v_{\pm , 1} v_{\pm , 2}$. In
type-I (type-II) DLNs, the product of Gaussian curvatures
\ba 
K_+ K_- = v_{+,1} v_{+,2} v_{-,1} v_{-,2} = 1 - \left( {\frac{\omega_1^2}{\lambda_1} }
+ { \frac{\omega^2}{\lambda_2^2}} \right)  
\ea 
are positive (negative). The final expression on the right follows from
Eq.\,(\ref{eq:Epm}).  As a result, all three methods, i.e. conic section
classification, sign inversion of band velocity, and band curvature analysis,
yield the same conclusion in regard to the type-I/II classification of DLNs. In
the type-II DLN, either the conduction band or the valence band has the
saddle-shaped band structure in the $(q_1,q_2 )$ space.

\section{Methods and \\ Electronic band structure}
\label{sec:methods}

\subsection{Calculation methods}
Having identified the physical and mathematical conditions to distinguish the
type-I and type-II DLNs, we now show that Na$_3$N under strain realizes a
type-II DLN. Before showing the first-principles results, we list the details
of computational methods. Our first-principles calculations are based on
density functional theory (DFT) in the Perdew--Zunger--type local density
approximation (LDA) \cite{Perdew81p5048}. The norm--conserving, optimized,
designed nonlocal pseudopotentials are generated using \textsc{opium}
\cite{Rappe90p1227}. The wave functions are expanded in plane--wave basis as
implemented in \textsc{quantum espresso} package \cite{giannozzi09p395502}. The
energy cutoff for the basis is set to 680\,eV. The atomic structure is fully
relaxed within the force threshold of 0.005\,eV/\AA\/ and the
8$\times$8$\times$8 Monkhorst-Pack grid \cite{Monkhorst76p5188} is used for
$k$-points sampling. The surface spectra are obtained by calculating the
surface Green's function for a semi-infinite geometry \cite{Sancho84p1205,
Sancho85p851}, using the Wannier Hamiltonian, generated using
\textsc{Wannier90} \cite{ZJWannier, Marzari97p12847, Mostofi08p685,
Souza01p035109}. For the Wannierization, the conduction and valence states are
initially projected to the $s$ orbitals of K and Na, and the $p$ orbitals of N,
for the wannierization. The lattice constant of Na$_3$N is calculated as $a$ =
4.67\,\AA. The lattice constants of strained Na$_3$N is calculated by fixing
$a$ and relaxing the other lattice parameter. In the case of 5\,\% of tensile
strain, we fixed $a$ to 4.91\,\AA, and relaxed $c$, resulting in 4.63\,\AA. The
Poisson ratio for the epitaxial strain is calculated as 0.2. In order to show
the robustness of our LDA results, the hybrid functional calculation for the
electronic band structure is performed using \textsc{vasp}
\cite{Kresse96p11169}, employing the Heyd-Scuseria-Ernzerhof (HSE06) scheme for
exchange-correlation potential \cite{2003JChPh.118.8207H}. Although SOC is
negligibly weak in Na$_3$N, the effect of SOC is considered in the Appendix B
for the sake of completeness of the study by using full-relativistic
pseudopotentials based on non-collinear spin calculations.

\subsection{Crystal structure and symmetries}
\begin{figure}[!ht]
\centering
\includegraphics[width=0.48\textwidth]{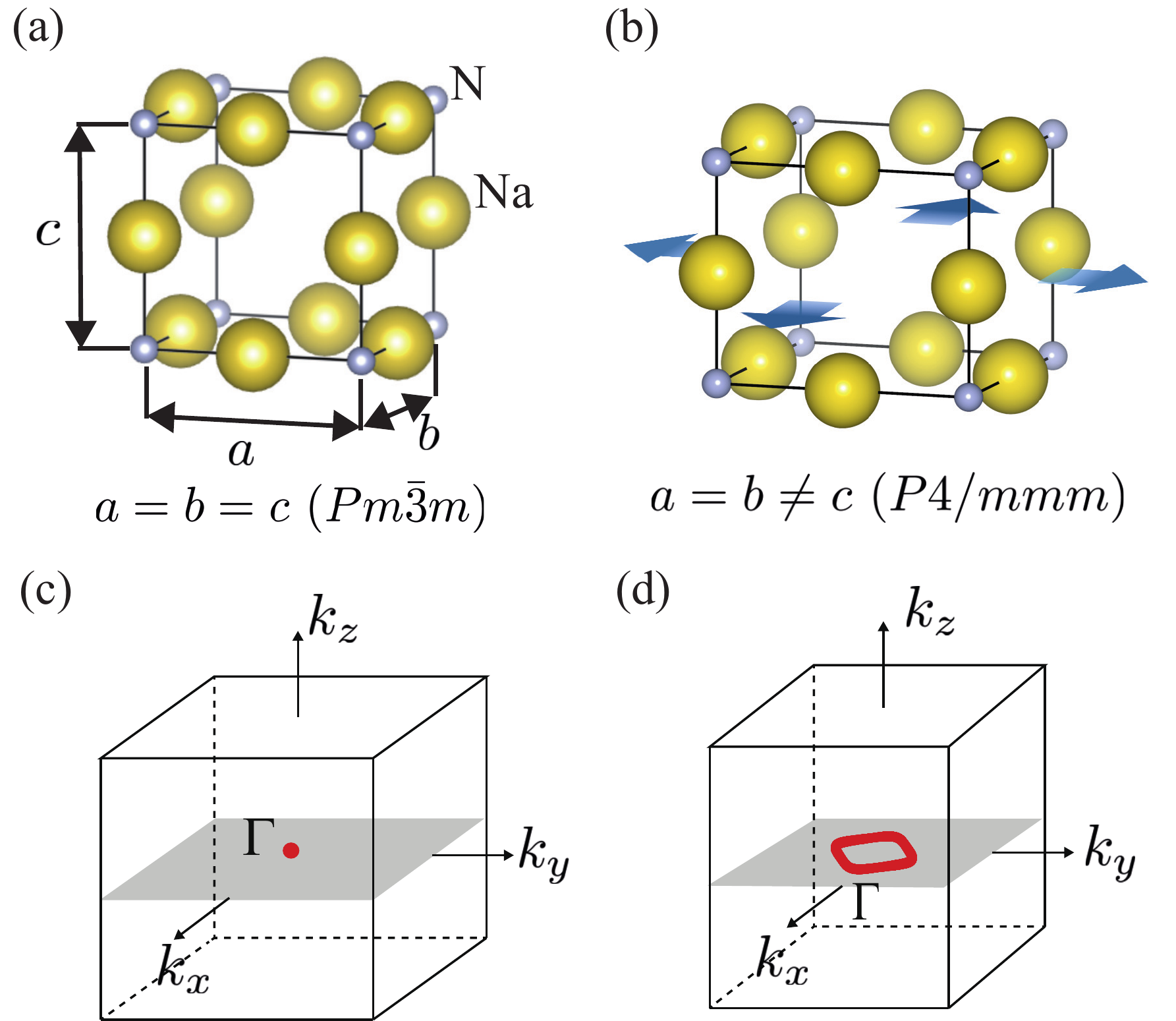} \\
\caption{Atomic structures of (a) pristine and (b) strained Na$_3$N. The blue
   arrows illustrate the direction of strain, which lowers the cubic
   $Pm\bar{3}m$ space group symmetry into the tetragonal $P4/mmm$ space group
   symmetry. The Brillouin zones (BZs) of (c) pristine
   Na$_3$N and (d) strained Na$_3$N. A nodal point of pristine Na$_3$N and a DLN of 
   strained Na$_3$N are colored by red.} \label{fig:fig2}
\end{figure}

Figure\,\ref{fig:fig2}(a) illustrates the primitive unit cell of Na$_3$N in
anti-ReO$_3$ structure, which comprises one N atom at the corner and three Na
atoms at the center of each edges. The crystalline symmetries belong to the
cubic space group Pm$\overline{3}$m (\#221), generated by inversion and three
rotations. Under an epitaxial strain shown in Fig.\,\ref{fig:fig2}(b), the
cubic crystalline symmetry is broken into the P4/mmm (\#123) tetragonal space
group symmetry. The P4/mmm space group is generated by inversion $\mathcal{P}$,
reflection $\mathcal{M}_x$, and the four-fold rotation $C_{4z}$.  Together with
these crystalline symmetries, time-reversal symmetry $\mathcal{T} =
\mathcal{K}$ plays an important role to provide topological protection of the
DLN. Here $\mathcal{K}$ is the complex conjugation operator.
Figure\,\ref{fig:fig2}(c) and (d) show the BZ of pristine Na$_3$N and strained
Na$_3$N, respectively. We find that the $\Gamma$ point in pristine Na$_3$N host
a triply-degenerate state (not counting the spin degeneracy), which evolves
into a DLN encircling $\Gamma$ lying on the $k_z = 0$ plane under the strain.
The evolution of the triply degenerate point to a DLN is detailed in subsection
\ref{subsection:DLN}.

\subsection{Electronic band structure}
\begin{figure}[!h]
\centering
\includegraphics[width=0.48\textwidth]{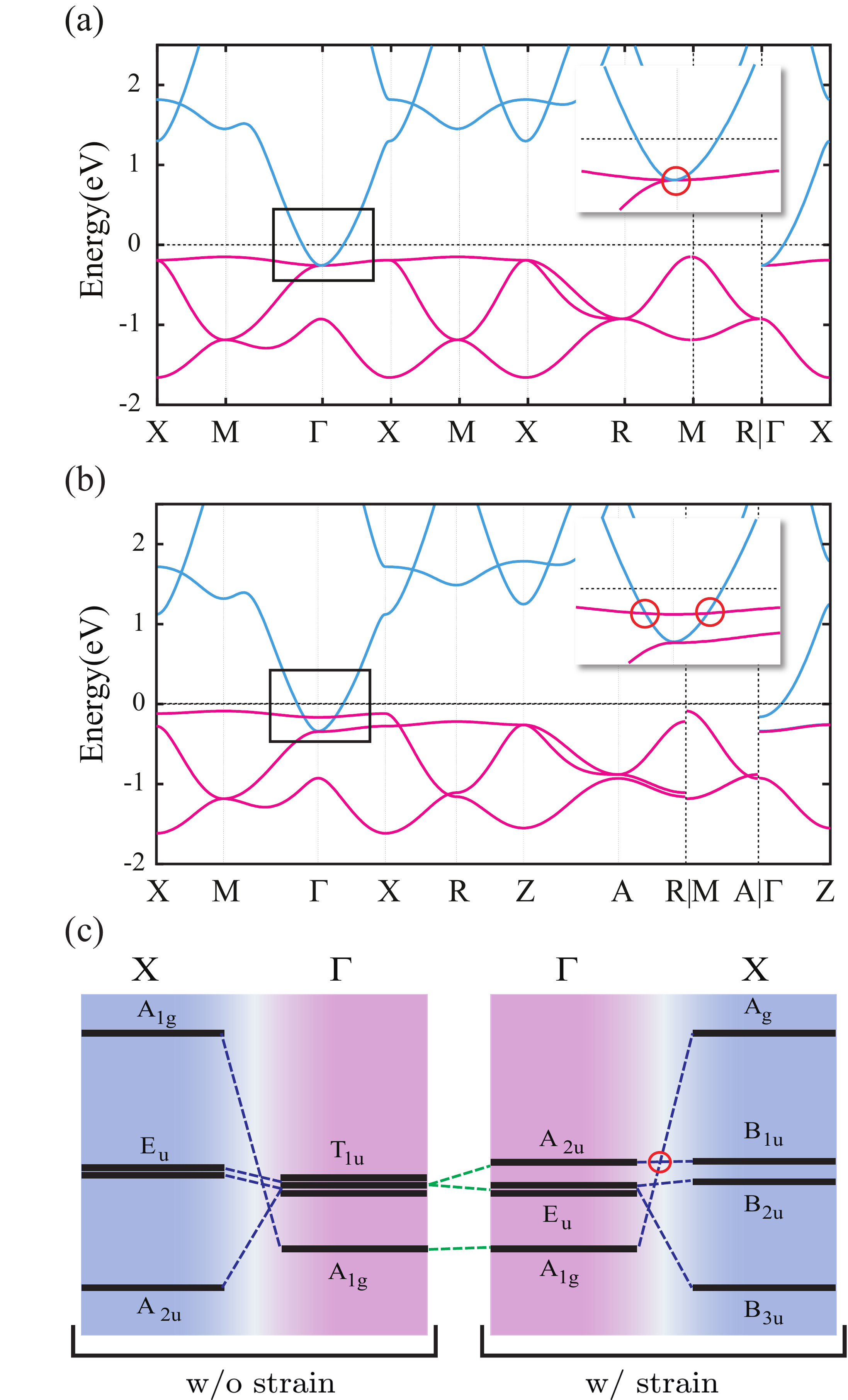} \\
\caption{Electronic band structures of (a) pristine and (b) 5\,\%-strained
   Na$_3$N. The band structures are drawn along high symmetry lines in the BZs
   of cubic and tetragonal unit cells, respectively. The conduction (valence)
   bands are represented by the blue (red) lines. The triply-degenerate point
   in Na$_3$N and doubly-degenerate points in strained Na$_3$N are indicated by
   red circles. (c) The diagram showing the evolution of energy ordering from
   $\Gamma$ to $X$ for pristine Na$_3$N and strained Na$_3$N. Lifting of triple
   degeneracy leads to the formation of a DLN.} \label{fig:fig3}
\end{figure}

Figure\,\ref{fig:fig3}(a) and (b) show the electronic band structures of
pristine and 5\,\% epitaxially tensile strained Na$_3$N obtained from our
first-principles calculations without spin-orbit coupling. It is clear that the
band structures of both cases are semimetallic, with the electron and hole
pockets forming near the $\Gamma$ and $M$ points, respectively. As it is well
known that LDA tends to underestimate the band gap (equally, overestimate the
band inversion), we also calculated the HSE06 band structure and find that the
semimetallic feature is reproduced \footnote{See Appendix A for the HSE06
results.}. Nonetheless, it is worth mentioning that, depending on the
exchange-correlation functionals, the band structure of Na$_3$N could be
semiconducting. Indeed, some literature \cite{VAJENINE2008450,Sommer2012}
claims that Na$_3$N is a semiconductor when a different exchange-correlation
energy functional is used. We discuss this further in Sec.\,\ref{sec:summary}.
We point out that the remainder of the discussion is based on our LDA and HSE06
results.

\section{Type-II DLN in strained Na$_3$N}
\label{sec:DLN}

As illustrated in Fig.\,\ref{fig:fig2}(d), an epitaxial tensile strain applied
to Na$_3$N engenders a DLN of the type-II class. In this section, we first show
that strained Na$_3$N hosts a DLN in momentum space.  The crossing points that
we found off the high symmetry  $\Gamma$ point in the band structure indicate
the presence of a DLN. From this indication, we identify the states forming the
DLN and clarify the geometry of the DLN in the BZ.  Next, we discuss the
topological protection by calculating $\mathbb{Z}_2$ topological invariants
that dictate the presence of DLNs \cite{Kim15p036806}. Finally, we show the
type-II nature of the DLN. For this purpose, we construct a $\v {k}\cdot\v {p}$
Hamiltonian and an explicit tight-binding model that reproduce the DFT band
structure, and apply the type-I/type-II criterion derived in
\ref{sec:backgrounds}, which confirms the type-II class. For further
demonstration of the type-II nature, we present the unconventional linear
dispersion occurring along principal axes from our DFT calculations. We also
calculate the numerical value of the principal curvatures, resulting in the
Gaussian curvatures with opposite sign, which reaffirms the type-II nature.
\subsection{DLN}
\label{subsection:DLN}

\begin{figure}
\centering
\includegraphics[width=0.48\textwidth]{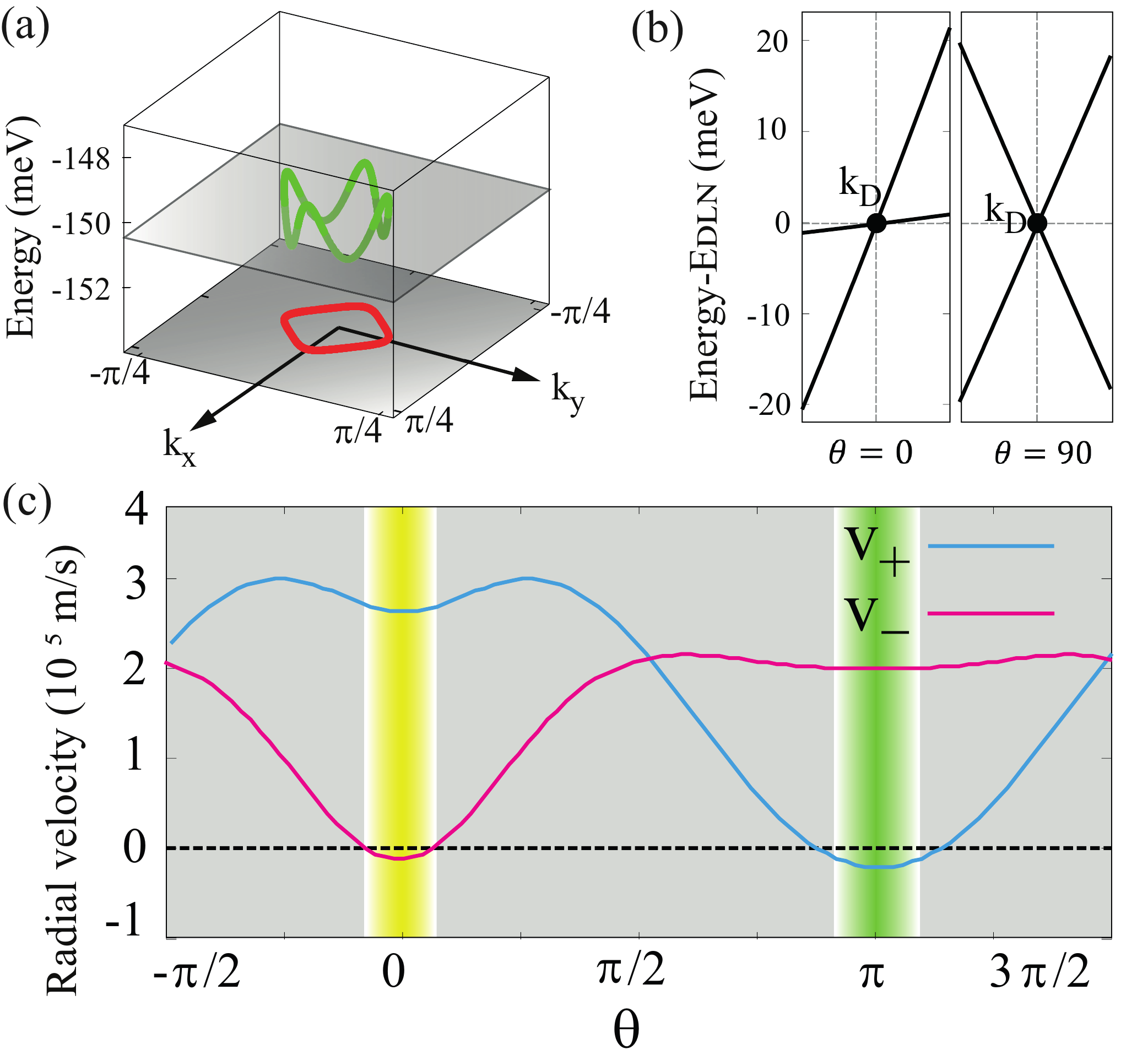} \\
\caption{(a) A DNL (red) in energy-momentum space. 
   A DLN is placed on the $k_z = 0$ plane, encircling the $\Gamma$
   point. The energy bands disperse along the DLN (green line) 
in the energy range of $\Delta E_{\textrm {DLN}}
   \approx 0.002$\,eV. (b) Energy dispersions  
 at a point (at $\v k_D$) of the type-II DLN drawn along 
 the $q_\rho$ (left panel) and $q_z$ (right panel) directions. 
   (c) The radial band velocities
   of the valence and conduction bands calculated as a function of
   the azimuthal angle $\theta$ on the $q_\rho$ - $q_z$ plane [see Fig.\,\ref{fig:fig1}(c)]. 
   The pink (blue) curve represents the velocity of the valence (conduction) band. 
   The valence (conduction)  band undergoes sign change outside (inside) of the DLN 
   across the critical angle
   $\theta \approx \pm 13.0^\circ $ ($\theta \approx 180^\circ \pm
   16.0^\circ $.) }
\label{fig:fig4}
\end{figure}

The low-energy electronic structure near the Fermi energy governs the
topological property of Na$_3$N. At $\Gamma$, we find that the valence band top
and the conduction band bottom are fused to form a triply-degenerate $T_{1u}$
state right below the Fermi level, indicated by a red circle in the inset of
Fig.\,\ref{fig:fig3}(a). In addition, a single $A_{1g}$ state resides 0.67\,eV
below the $T_{1u}$ state at $\Gamma$. An epitaxial strain gives rise to a
tetragonal crystal field, which splits the triply-degenerate $T_{1u}$ state
into a doubly-degenerate $E_{u}$ state and a single $A_{2u}$ state. Tensile
(compressive) strain raises (lowers) the energy of the $A_{2u}$ state with
respect to the doubly-degenerate $E_{u}$ state at $\Gamma$.  The
energy-lowering of $A_{2u}$ under a tensile strain enables the crossing between
$A_{2u}$ and $A_{1g}$  bands. Notably, the crossing does not happen when
applying the compressive strain that raises the energy of $A_{2u}$ state. The
inversion of the $A_{1g}$ and $A_{2u}$ states is responsible for the occurrence
of a DLN in strained Na$_3$N, which occurs along the high-symmetry lines
$\Gamma-M$ or $\Gamma-X$. This results in band-crossing points indicated by red
circles in Fig.\,\ref{fig:fig3}(b) and (c).  Careful inspection of the band
structure in the entire BZ allows us to find a single DLN encircling $\Gamma$
that lies on the $k_z = 0 $ plane as shown in Fig.\,\ref{fig:fig4}(a). The DLN
is located at $E_{\rm DLN} \approx - 0.15$\,eV below the Fermi energy, with
dispersion in the energy range of $\Delta E_{\rm DLN} \approx 0.002$\,eV.

\subsection{Topological protection}
 The protection of the DLN is two-fold. First, lying on the
$k_z = 0$ mirror-invariant plane, the DLN is protected by $\mathcal{M}_z$ with
the conduction and valence bands having different mirror eigenvalues $\pm1$. In
addition, the DLN is also topologically protected by inversion $\mathcal{P}$
and time-reversal $\mathcal{T}$ symmetries \cite{Kim15p036806, Yu15p036807}.
Judging from the group representation of the Bloch states forming the DLN, we
find the band inversion at $\Gamma$ between $A_{2u}$ and $A_{1g}$ states, which are odd and even parity eigenstates of inversion ${\cal P}$,
respectively. These two states are also mirror eigenstates having opposite
parities, leading to the mirror-protected band crossing on the mirror-invariant plane $k_z=0$. The presence and geometric shape of a DLN in
momentum space is dictated from the $\mathbb{Z}_2$ topological indices
$(\nu_0;\nu_1\nu_2\nu_3)$ as discussed in Ref.\,\cite{Kim15p036806}. We find
$(\nu_0;\nu_1\nu_2\nu_3) = (1;000)$ for strained Na$_3$N from parity analysis at the time-reversal invariant momenta (TRIMs) $(\Gamma, 2X, M, 2R, A, Z)$, which agree with the existence and shape of the DLN. In detail, $(\nu_0;\nu_1\nu_2\nu_3) = (1;000)$ enforces an odd number of DLNs to thread the half $\mathcal{T}$-invariant plane containing $\Gamma$. This is fulfilled by the
DLN encircling the $\Gamma$ lying on the $k_z = 0$ plane.

\subsection{$\v {k}\cdot\v {p}$ analysis and type-II nature}
In this section, we construct the $\v {k}\cdot\v {p}$ Hamiltonian that
describes the DLN, and confirm its type-II nature. The $\v {k}\cdot\v {p}$
Hamiltonian near $\Gamma$ can be constructed such that it respects the $O_h^1$
little group symmetries of $\Gamma$ on the basis $\binom{1}{0} = A_{2u}$ and
$\binom{0}{1} = A_{1g}$
\begin{eqnarray}
H & = &  [ a_1(k_\rho^2-k_D^2) + a_2 k_z^2 ]  + v k_z\sigma_y  \nonumber \\
& + & [ b_1(k_\rho^2-k_D^2) + b_2 k_z^2 ] \sigma_z .
\label{eq:eq6}
\end{eqnarray}
Here $k_{\rho} = \sqrt{k_x^2 + k_y^2}$, and $\sigma_j$'s are the Pauli matrices
in (A$_{2u}$, A$_{1g}$) basis, and $a_1$, $a_2$, $b_1$, and $b_2$ are
constants. The DLN forms along a circle $k_\rho = k_D$ on the $k_z = 0$ plane.
The first term of Eq.\,(\ref{eq:eq6}) induces tilting that enables the type-II
nature of the DLN. The best match to the DFT band structure in the vicinity of
the $\Gamma$ point for pristine and 5\,\%-strained Na$_3$N is found when using
the parameters listed in Table\,\ref{table:kp} \footnote{See Appendix C more
parameter sets for 1\%, 2\%, 3\%, and 4\%-strained cases.}. In the pristine
case, where $a_1 =  a_2$, $b_1 = b_2$, $v  = 0$, and $k_D = 0$, the energy
state are degenerate  at $\Gamma$ [$(k_\rho, k_z=(0, 0)$], exhibiting isotropic
band structure originated from the cubic symmetry. The tensile strain reduces
the cubic symmetry to tetragonal symmetry, described by non-zero values for the
velocity $v$ and the radius of the DLN $k_D$, as well as the 	difference
between the values of $a_1 (b_1)$ and  $a_2 (b_2)$.

\begin{table}[tp]
\caption{Parameters of the $\v {k}\cdot\v {p}$ Hamiltonian that best match the
   DFT band structures of pristine and 5\,\%-strained Na$_3$N. Here, $k_{\rho}$
   and $k_{z}$ are in $2\pi/a$ and $2\pi/c$ units, respectively, and the
   parameters are in eV unit. }
\label{table:parameter}
\centering
\begin{tabular}{c c c}
\hline
Parameters & Pristine  & 5\,\%-strained\\
\hline
$a_1$         & 5.00 & 5.00  \\
$b_1$         & 4.60 & 4.60  \\
$a_2$         & 5.00 & 4.00  \\
$b_2$         & 4.60  & 3.80  \\
$v$             & 0.00 & 1.05  \\
$k_D$        & 0.00  & 0.14  \\
\hline
\end{tabular}
\label{table:kp}
\end{table}

As previously pointed out in Sec.\,\ref{sec:backgrounds}, the $\v k \cdot \v p$
Hamiltonian can be expanded in $\v q = \v k - \v k_D$ in the vicinity of a
point on the DLN $\v k_D$

\begin{eqnarray}
H = 2 a_1 k_D q_\rho + v q_z \sigma_y + 2 b_1 k_D q_\rho \sigma_z ,
\label{eq:eq7}
\end{eqnarray}
which is of the same form as Eq.\,(\ref{eq:eq2}) with the principal axes
$q_\rho$ and $q_z$. The type-II criterion previously worked out in Sec.
\ref{sec:backgrounds} is simplified as
\ba a_1^2 > b_1^2
\label{eq:cri}, \ea
which is consistent with the previously proposed type-I/type-II classification
of DLNs \cite{Li2017}. As shown in Table\,\ref{table:kp}, 5\,\%-strained
Na$_3$N indeed  satisfies Eq.\,(\ref{eq:cri}), thus confirming the type-II DLN
semimetal phase.

The type-II nature is further illustrated from the shape of Dirac cones.
Figure\,\ref{fig:fig4}(b) illustrates the band structures drawn along the
$q_\rho$ ($\theta =0^\circ$) and $q_z$ ($\theta = 90^\circ$) directions, which
are the principal axes of the $\v {k}\cdot\v {p}$ Hamiltonian.  We find that
the unconventional (left panel) and conventional (right panel) linear
dispersions coexist in the vicinity of the Dirac point, which is a
characteristic of the unconventional type-II DLN. Figure\,\ref{fig:fig4}(c)
shows the radial band velocities $v_\pm$ around the DLN as a function of
$\theta=[0,2\pi]$ calculated from the DFT band structures. Conventional
(unconventional) linear dispersion resides in grey (yellow and green) region,
resulting in $v_+ v_- > 0$ ($v_+ v_- < 0$), which proves the sign change in the
band velocities. Additionally, it is manifested that either the conduction or
the valence band is a saddle-shaped band undergoing the sign flip of the radial
band velocity, resulting in $K_+ K_- < 0$. This is a characteristic of the
radial band velocities only present in a type-II DLN semimetal. In a type-I DLN
semimetal, $v_\pm$ should be positive for the entire range of
$\theta=[0,2\pi]$, resulting in $K_+ K_- > 0$.

\section{Tight-binding model analysis}
\label{sec:tb}
The tight-binding model is a useful way to develop the low-energy effective
theory for the DLN. We construct it using the basis of $p$-orbitals
of the N atom and the $s$-orbital of Na. The basis orbitals are thus (N $p_x$,
N $p_y$, N $p_z$, Na-$x$ $s$, Na-$y$ $s$, Na-$z$ $s$), where Na-$x$ is the Na
atom at $(a/2, 0, 0)$, Na-$y$ at $(0, a/2, 0)$, and Na-$z$ at $(0, 0, c/2)$,
respectively. We arrive at the tight-binding Hamiltonian written as a block form

\ba \mathcal{H}  = \begin{pmatrix} {\cal H}_{\rm N} & {\cal H}_{\rm N\!-\!Na}
\\ H_{\rm N\!-\!Na}^\dag & {\cal H}_{\rm Na} \end{pmatrix} .  \ea The
$3\times3$ submatrices ${\cal H}_{\rm N}$, ${\cal H}_{\rm Na}$ concern the
three $p$-orbitals of N and the three inequivalent $s$-orbitals of Na,
respectively. ${\cal H}_{\rm Na-N}$ is their hybridization. In detail,

\ba {\cal H}_{\rm N} = \begin{pmatrix}
\begin{matrix}  E_{{\rm N} , p}  \\
- 2  t_{3pp\sigma\parallel} c_x \\
+ 2   t_{3pp\pi\parallel}  c_y  \\
+ 2   t_{3pp\pi\perp }  c_z
\end{matrix}   & 0 & 0 \\ 0 &
\begin{matrix} E_{{\rm N} , p}  \\
+ 2    t_{3pp\pi\parallel} c_x\\
- 2   t_{3pp\sigma\parallel} c_y\\
+ 2    t_{3pp\pi\perp} c_z
\end{matrix} & 0 \\ 0 & 0 &
\begin{matrix} E_{{\rm N} , p} \\
+ 2   t_{3pp\pi\parallel}  c_x\\
+ 2    t_{3pp\pi\parallel} c_y\\
- 2   t_{3pp\sigma\perp} c_z
\end{matrix}
\end{pmatrix} \nonumber \ea

\ba {\cal H}_{\rm Na} = \begin{pmatrix}
E_{\mathrm{Na},s}  &
4 t_{2ss \parallel}  c_{x2} c_{y2} &
4 t_{2ss \perp} c_{x2} c_{z2}  \\
4 t_{2ss \parallel}  c_{x2} c_{y2} &
E_{\mathrm{Na},s} &
4 t_{2ss \perp}  c_{y2} c_{z2} \\
4 t_{2ss \perp} c_{x2} c_{z2}  &
4 t_{2ss \perp} c_{y2} c_{z2}  &
E_{\mathrm{Na},s}
\end{pmatrix} \nonumber \ea

\ba
{\cal H}_{\rm N\!-\!Na} = \begin{pmatrix} 2 i t_{sp\sigma\parallel}   s_{x2}  & 0 & 0 \\
0 & 2 i t_{sp\sigma\parallel}   s_{y2}  & 0 \\
0 & 0 & 2 i t_{sp\sigma\perp}   s_{z2}  \end{pmatrix} . 
\ea
Here, the abbreviations are used for $c_{x(y)} = \cos k_{x(y)}$, $c_{z} = \cos
k_{z} $, $c_{x(y) 2} = \cos(k_{x(y)}/2)$, $c_{z2} = \cos(k_z/2)$, $s_{x(y) 2} =
\sin(k_{x(y)}/2)$, $s_{z2} = \sin(k_z/2)$. The on-site energies of N $p$ and Na
$s$ orbitals are denoted as $E_{\rm N,p}$ and $E_{\rm Na,s}$, respectively,
which are set to the same values both in pristine and strained cases.

We consider the hopping between the basis orbitals up to third nearest
neighbors in the pristine and strained crystal structure. For the nearest
neighbor hopping, the hopping integral between the $p_{x (y,z)}$ orbital at the
origin and the $s$-orbital of the Na-$x (y,z)$ atom is denoted as
$t_{sp\sigma,x (y,z)}$. In the pristine case, the cubic symmetry enforces
$t_{sp\sigma,x}=t_{sp\sigma,y}=t_{sp\sigma,z} = t_{sp\sigma}$. In the strained
case, the lowered symmetry differentiate $t_{sp\sigma,z}$ from $t_{sp\sigma,x}
= t_{sp\sigma,y}$. We denote the $t_{sp\sigma,x (y)}$ as
$t_{sp\sigma\parallel}$ and $t_{sp\sigma,z}$ as $t_{sp\sigma\perp}$,
respectively. The second-nearest neighbor hopping takes place between Na-$x$
and Na-$y$, or between equivalent Na atoms. The corresponding hopping integrals
are denoted as $t_{2ss \parallel}$ for Na-$x$ $s$ to Na-$y$ $s$, and $t_{2ss ,
\perp}$ for Na-$x$ $s$ to Na-$z$ $s$, or Na-$y$ $s$ to Na-$z$ $s$. One can
expect $t_{2ss \parallel} = t_{2ss \perp}$ only for the pristine cubic crystal
of Na$_3$N. The third-nearest neighbor hopping takes place between the $p$
orbitals of N atoms in the adjacent unit cells [distanced by $\pm R_{x
(y,z)}$], parameterized by the hopping integrals $t_{3 pp,x (y,z)}$.
Considering $\pi$ and $\sigma$ symmetry between $p$ orbitals, we obtain the
constraints for the pristine and strained cases. In the pristine case, there is
no difference among $x$, $y$, and $z$-directions. Therefore, there are only two
types of distinct hopping integral $t_{3pp\sigma}$ and $t_{3pp\pi}$. In
contrast, there are four different third neighbor hopping integral
$t_{3pp\sigma\parallel}$, $t_{3pp\sigma\perp}$, $t_{3pp\pi\parallel}$, and
$t_{3pp\pi\perp}$ in the strained case. In the tetragonal-symmetric case,
$t_{3pp\sigma (\pi)\parallel}$ represents $t_{3pp\sigma (\pi), x}$ and
$t_{3pp\sigma (\pi), y}$. On the other hand, $t_{3pp\sigma (\pi), \perp}$
represents $t_{3pp\sigma (\pi), z}$ The on-site energies and hopping integrals
for pristine (strained) cases are listed in Table\,\ref{table:pristine.values} (\ref{table:strained.values}). A negative value of the hopping parameter
$t_{3pp\pi \parallel}$ is responsible for the DLN belonging to the type-II
class. A type-I DLN occurs for positive $t_{3pp\pi \parallel}$.

\begin{figure}[!htb]
\includegraphics[width=0.48\textwidth]{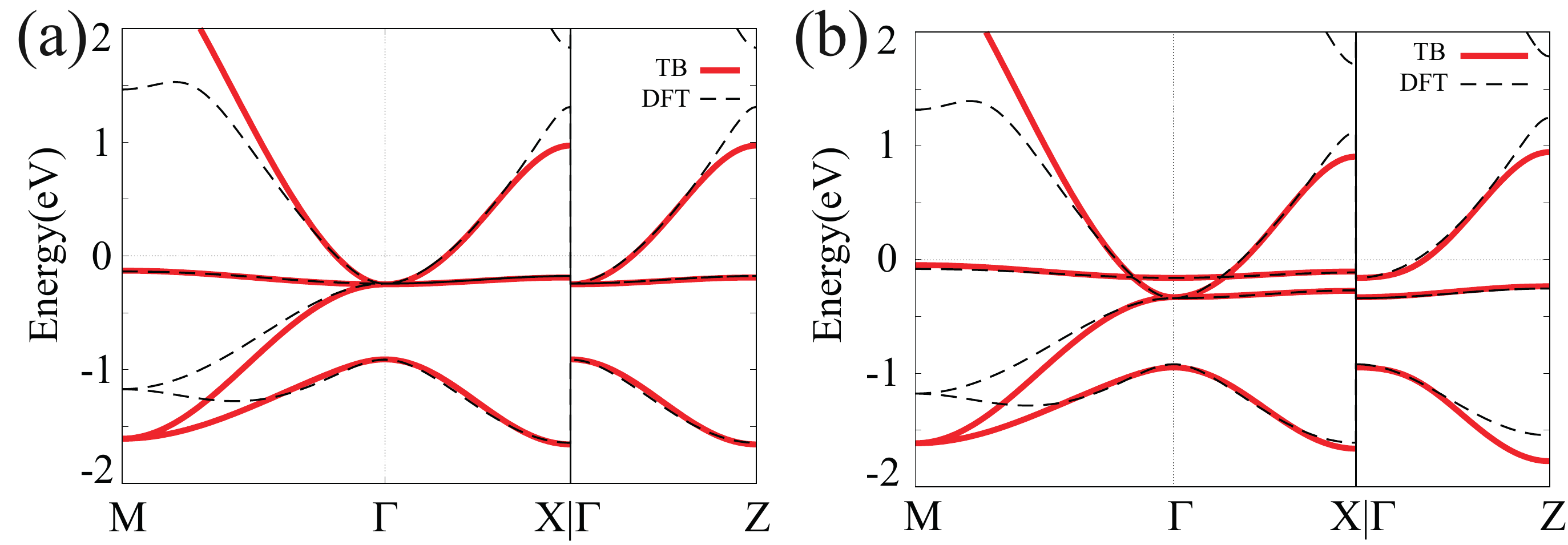}
\caption{Tight-binding and DFT band structures for (a) pristine and (b)
   5\,\%-strained Na$_3$N.}
\label{fig:TB-fit}
\end{figure}

Close fits to both pristine and strained DFT bands are achieved in
Fig.\,\ref{fig:TB-fit} using the tight-binding parameters in
Table\,\ref{table:tightBindingHopping}. Near the $\Gamma$ point, the DLN occurs.
The tight-binding model corroborates the following DFT results. The line node
is formed via the band inversion between the odd-parity eigenstate composed of
N $p_z$ orbital, and the even-parity eigenstate composed of three $s$-orbitals
from Na-$x$, Na-$y$, and Na-$z$. In both cases, the even parity state is below
the odd parity eigenstate at $\Gamma$, while the even parity state is above the
odd parity eigenstates in other seven TRIM points. The strain opens the band
gap between $n_{occ}=3$ and $n_{occ}=4$ states at the eight TRIM points,
resulting in non-trivial ${\cal Z}_2$ topological invariant. Based on the
tight-binding parameters of Table\,\ref{table:strained.values} we also
reproduced the same ${\cal Z}_2$ topological invariants as the DFT results.

\begin{table}[tp]
\caption{On-site energies and hopping integrals for pristine Na$_3$N. }
\label{table:tightBindingHopping}
\centering
\begin{tabular}{c c}
\hline
On-site energy & Value (eV) \\
\hline
$E_{\mathrm{N},p}$                & -2.45 \\
$E_{\mathrm{Na},s}$              & +0.85 \\
\hline
Hopping integral &  \\
\hline
$t_{sp\sigma}$ & -1.00 \\
$t_{2ss}$          & -0.47 \\
$t_{3pp\sigma}$ &  -0.13 \\
$t_{3pp\pi}$   & -0.015 \\

\hline
\end{tabular}
\label{table:pristine.values}
\end{table}

\begin{table}[tp]
\caption{On-site energies and hopping integrals for 5\,\%-strained Na$_3$N. }
\label{table:tightBindingHopping_strained}
\centering
\begin{tabular}{c c}
\hline
On-site energy & Value (eV) \\
\hline
$E_{\mathrm{N},p}$                & -2.45 \\
$E_{\mathrm{Na},s}$              & +0.85 \\
\hline
Hopping integral & \\
\hline
$t_{sp\sigma\parallel}$   &  -0.95 \\
$t_{sp\sigma\perp}$       &  -0.95 \\
$t_{2ss\parallel}$            &  -0.46 \\
$t_{2ss\perp}$                &  -0.47 \\
$t_{3pp\sigma\parallel}$ &  -0.13 \\
$t_{3pp\sigma\perp}$     &  -0.205 \\
$t_{3pp\pi\parallel}$       & -0.014 \\
$t_{3pp\pi\parallel}$       & -0.025 \\

\hline
\end{tabular}
\label{table:strained.values}
\end{table}

\section{Physical manifestations}
\label{sec:physical}
In this section, we discuss feasible ways to detect the type-II DNL character
of strained Na$_3$N. Two physical manifestations are suggested: topological
surface states and optical conductivity that we discuss in the following
subsections. We suggest that the two can help clarify the topological and
type-II nature of the DLN hosted in strained Na$_3$N, respectively.

\subsection{Topological surface states}
\begin{figure}[!h]
\centering
\includegraphics[width=0.48\textwidth]{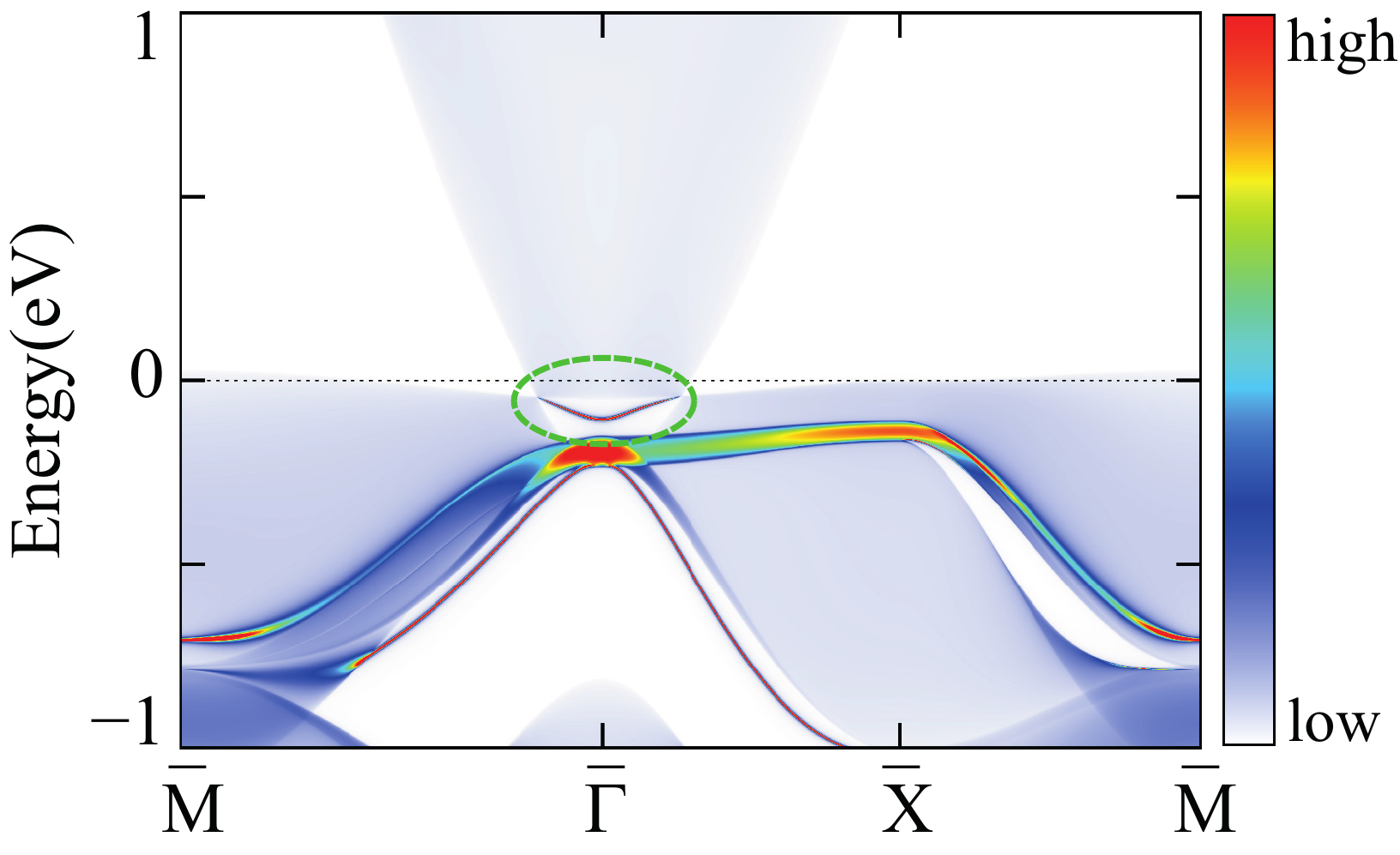} \\
\caption{Surface band structure of 5\,\%-strained Na$_3$N calculated from the
   maximally localized Wannier Hamiltonian. Topological surface states appear
   inside the bulk gap near the $\Gamma$ point as a bright branch inside a
   green circle. The color scheme shows the density of states at a given energy
   and $k$-points. Red (blue) means high (low) density of the surface projected
   states.}
\label{fig:surf.s}
\end{figure}
A topologically protected DLN features drumhead-like surface states
\cite{Burkov2011, Kim15p036806, Yu15p036807, Weng15p045108}. We confirm this
topological characteristic of the surface energy spectra using the surface
Green's function. Figure\,\ref{fig:surf.s} shows the resultant surface band
structure of the (001) surface for the semi-infinite slab of Na$_3$N, drawn
along projected high-symmetry lines following
$\bar{M}-\bar{\Gamma}-\bar{X}-\bar{M}$ path of the square lattice. As the bulk
DLN is parallel to the (001) surface, the interior region of the DLN is
projected to a finite area of the surface BZ near $\Gamma$. The region shaded
by a bluish color in Fig.\,\ref{fig:surf.s} shows the projected bulk states. The
type-II nature is revealed in the unconventional (tilted) linear dispersion
appearing on $\bar{M}-\bar{\Gamma}$ and $\bar{\Gamma}-\bar{X}$, as shown in
Fig.\,\ref{fig:surf.s}.

The high-intensity branch connecting the two crossing points is the topological
surface states. They appear in the interior region of the projected DLN
enclosing $\bar{\Gamma}$, indicated by the green circle. This explicitly proves
the topological nature of strained Na$_3$N. As discussed in the
Ref.\,\cite{Kim15p036806}, the curvature of the surface states is in part
determined by the harmonic average of the curvatures of the conduction and
valence bands. Our DFT calculations also feature this, yet due to the same sign
of the curvature for the conduction and valence bands, the topological surface
states appear more dispersive than nearly flat surface bands of a type-I DLN.
This is another characteristic feature of a type-II DLN semimetal captured in
the surface energy spectrum. We also note that there is high intensity
contribution  from the bulk states to the surface energy spectrum, originated
from non-dispersive bulk bands along the $z$ direction, comprising mainly the N
$p_x, p_y$ orbitals. We suggest that these non-topological states should be
well-separated from the topological drumhead-like states when the strain is
applied beyond the 4\,\% of tensile strain, making clear distinction between
the topological surface states from the bulk trivial states.

\subsection{Optical conductivity}
\label{subsec:opcd}
\begin{figure}[!htb]
\includegraphics[width=0.48\textwidth]{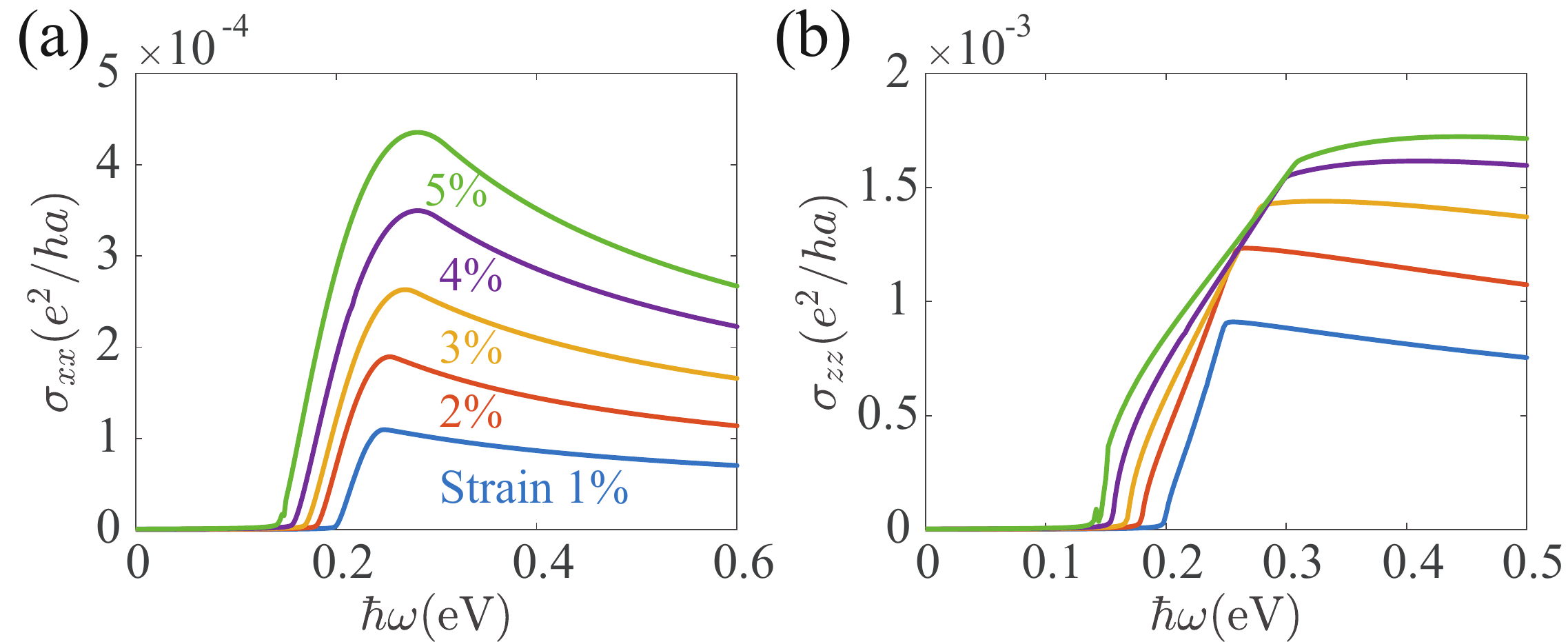}
\caption{Interband optical conductivities of Na$_3$N as a function of an
   epitaxial tensile strain. Both (a) $\sigma_{xx}$ and (b) $\sigma_{zz}$ show
   a sudden jump, which occurs near the energy of the DLN. The optical
   conductivity vanishes for the prinstine case because the interband
   transitions are forbidden.}
\label{fig:opcd}
\end{figure}
The optical conductivity of DLN semimetals has been studied in the previous
literature and is known to exhibit a flat behavior at low frequencies below the
nodal ring energy scale \cite{Ahn2017,Carbotte2017}. In this subsection, we
investigate the optical responses of Na$_3$N for both the pristine and strained
cases. We demonstrate that the strain results in a qualitative change
in the optical conductivity, suggesting that the optical signal of Na$_3$N
could serve as an experimental footprint for the ocurrence of the DLN under
strain.

To obtain the optical conductivity, we evaluate the Kubo formula using the
two-band effective model in Eq.\,(\ref{eq:eq6}). In the linear response and
clean limit, the Kubo formula for the optical conductivity \cite{Mahan2013} is
written as

\begin{equation}
\begin{split}
\label{eq:KuboFormula}
\sigma_{ij}(\omega)
&=- \frac{ie^2}{\hbar} \sum_{s,s'} \int \frac{d^3 k}{(2\pi)^3} \frac{f_{s, \bm{k}}-f_{s',\bm{k}}}{\varepsilon_{s,\bm{k}}-\varepsilon_{s',\bm{k}}} \\
&\times
\frac{M^{ss'}_i(\bm k)M^{s's}_j(\bm k)}{\hbar\omega+\varepsilon_{s,\bm{k}}-\varepsilon_{s',\bm{k}}+i0^+},
\end{split}
\end{equation}
where $i,j=x,y,z$, $\varepsilon_{s,\bm{k}}$ and
$f_{s,\bm{k}}=1/[1+e^{(\varepsilon_{s,\bm{k}}-\mu)/k_{\rm B}T}]$ are the
eigenenergy and the Fermi distribution function for the band index $s=\pm$ and
wave vector $\bm{k}$, respectively, $\mu$ is the chemical potential and
$M^{ss'}_i(\bm k)=\langle{s,\bm{k}}|\hbar\hat{v}_i |{s',\bm{k}}\rangle$ with
the velocity operator $\hat{v}_i=\frac{1}{\hbar}\frac{\partial
\hat{H}}{\partial  k_i}$.

We first consider the pristine case. By applying the pristine parameters
in Table\,\ref{table:kp} to Eq.(\ref{eq:eq6}), we obtain the effective
Hamiltonian $H=a_1\left|\bm{k}\right|^2+b_1\left|\bm{k}\right|^2\sigma_z$. The
energy eigenstates and the velocity operator can be represented as the eigenstates of
$\sigma_z$ and $\hat{v}_i=\frac{2k_i}{\hbar}  (a_1 + b_1 \sigma_z)$,
respectively. Using these, it is readily shown that the matrix element
$M^{ss'}_i(\bm k)\propto\delta_{ss'}$, and thus the interband transitions are
forbidden in the pristine case.
In contrast, in the strained case, the interband transitions are allowed,
thus resulting in the non-zero optical response.
The non-zero velocity term [$v k_z\sigma_y$ in Eq.(\ref{eq:eq6})] 
makes the wavefunctions $\v k$-dependent, leading to the
non-vanishing matrix element for the interband transitions.
We have calculated the interband optical conductivity as a function of strain; 
results for 1\,\% to 5\,\% are summarized in Fig\,\ref{fig:opcd} \footnote{See
Appendix \ref{table:kp2}
for the $\v k \cdot \v p$ parameters that we used to calculate the optical
conductivities}. Notably, both $\sigma_{xx}$ and $\sigma_{zz}$ exhibit a sudden
increase at around $\hbar\omega = E_\mathrm{DLN}$ for a given strain, which
corresponds to the size of the optical gap arising  from the Pauli blocking.
Such a qualitative change upon straining the material, i.e., a sudden rise of
the optical conductivity, is a key signature of the emergence of the Dirac line
node induced by strain. 

\section{Discussion and Summary}
\label{sec:summary}

In summary, we characterized a type-II topological DLN semimetal and proposed
strained Na$_3$N as its material realization.  We showed that type-I/type-II
DLNs can be classified in equivalent manner by the mathematical formalism
governing any one of the three features: Fermi surface geometry, sign inversion
of the band velocity, or the band curvature.  We believe that these connections
provide a fundamental and clear picture for the type-I/type-II classification
of DLNs. In addition, the Type-II condition represented in terms of the band
velocity should provide a computationally convenient way to determine the types
of DLNs. Furthermore, our extensive DFT calculations predict that a type-II DLN
semimetal should be realized in Na$_3$N under epitaxial strain. We propose the
drumhead surface states spectrum and optical conductivity as two key physical
manifestations indicating the existence of a type-II DLN in strained Na$_3$N.
In particular, the optical response is expected to undergo a sudden jump under
the strain due to a creation of a DLN, while the interband transition in
pristine Na$_3$N is suppressed by the selection rule. This feature should serve
as an experimental evidence of the type-II DLN semimetal phase hosted in
strained Na$_3$N.

Encouragingly, the synthesis of pristine Na$_3$N has been reported in the
literature \cite{Fischer2002, Niewa2002, Vajenine2007}. However to the best of
our knowledge, the literature disagrees with our DFT calculation. For example,
an optical response of Na$_3$N reported in Ref.\,\cite{VAJENINE2008450} for
visible light and near IR spectra results in a sizable optical band gap of 1.6
- 2.0\,eV. Also, first-principles calculations based on self-interaction
correction (SIC) and G$_0$W$_0$ methods support the experiment
\cite{Sommer2012}. Clearly, they stand in contrast with our calculation based
on LDA and HSE. On the other hand, the band gap calculation based on the HSE06
hybrid functional, which reproduce experimental band gaps with a high degree of
accuracy in some systems \cite{Lucero2011, Paier2009, PhysRevB.82.205212,
Gillen2013, Moses2011}, also shows the metallic behavior in line with the LDA
and GGA calculations (see Appendix A).

As described in Sec.\,\ref{sec:DLN}, the inversion of $s$ and $p$ bands at the
$\Gamma$ point is a key ingredient for the formation of a DLN as well as the
band gap. The band inversion is captured by the LDA or HSE06 calculation, but
apparently not by the previous SIC or G$_0$W$_0$ calculations. To better
understand the origin of this discrepancy, we made tight-binding fits to band
structures of HSE06, SIC, and G$_0$W$_0$ as shown in the Appendix A. Compared
to the estimation by the LDA results, other calculations gave an increase in
the on-site energy of Na $s$-orbitals $E_{\rm Na,s}$ given by +0.65, +1.10, and
2.25\,eV, respectively.  The dramatic increase in $E_{\rm Na, s}$ in the case
of SIC and G$_0$W$_0$ is presumably responsible for the lack of band inversion
at the $\Gamma$ point, as well as the consequent absence of a DLN.
Tight-binding models deduced from fits of SIC and G$_0$W$_0$ bands gave the
trivial topological number, as expected.

We believe that the metallicity of Na$_3$N yet requires further confirmation
via accurate band gap measurements, such as absorption spectra with low-energy
photon energies or transport experiments. The previous experiment was performed
using Na3N powders that could potentially contain excessive Na, as mentioned in
the study \cite{VAJENINE2008450}. Therefore, existing discrepancy regarding the
metallicity of Na$_3$N is a source for pursing careful experimental
verification of electronic properties of this material, given our finding of
the topological nodal semi-metallic behavior with novel velocity dispersion
under the strain.

\begin{acknowledgments}
D. K. was supported by Samsung Science and Technology Foundation
   (SSTF-BA1701-07) and Basic Science Research Program through the National
   Research Foundation of Korea (NRF) funded by the Ministry of Education
   (NRF-2018R1A6A3A11044335). S.A. and H.M. were supported by the NRF grant
   funded by the Korea government (MSIT) (No. 2018R1A2B6007837) and
   Creative-Pioneering Researchers Program through Seoul National University.
   J. H. H. was supported by Samsung Science and Technology Foundation under
   Project Number SSTF-BA1701-07. Y.K. was supported from Institute for Basic
   Science (IBS-R011-D1) and the NRF grant funded by the Korea government
   (MSIP; Ministry of Science, ICT $\&$ Future Planning) (No.
   S-2017-0661-000).  The computational calculations were performed using the
   resource of Korea institute of Science and technology information (KISTI).
\end{acknowledgments}

\bibliography{refs}

\setcounter{equation}{0}
\setcounter{figure}{0}
\setcounter{table}{0}
\renewcommand{\theequation}{S\arabic{equation}}
\renewcommand{\thefigure}{S\arabic{figure}}
\renewcommand{\thetable}{S\arabic{table}}
\section*{Appendix A: Band structure calculation with other type of
exchange-correlation functionals}

\begin{figure}[!h]
\centering
\includegraphics[width=0.48\textwidth]{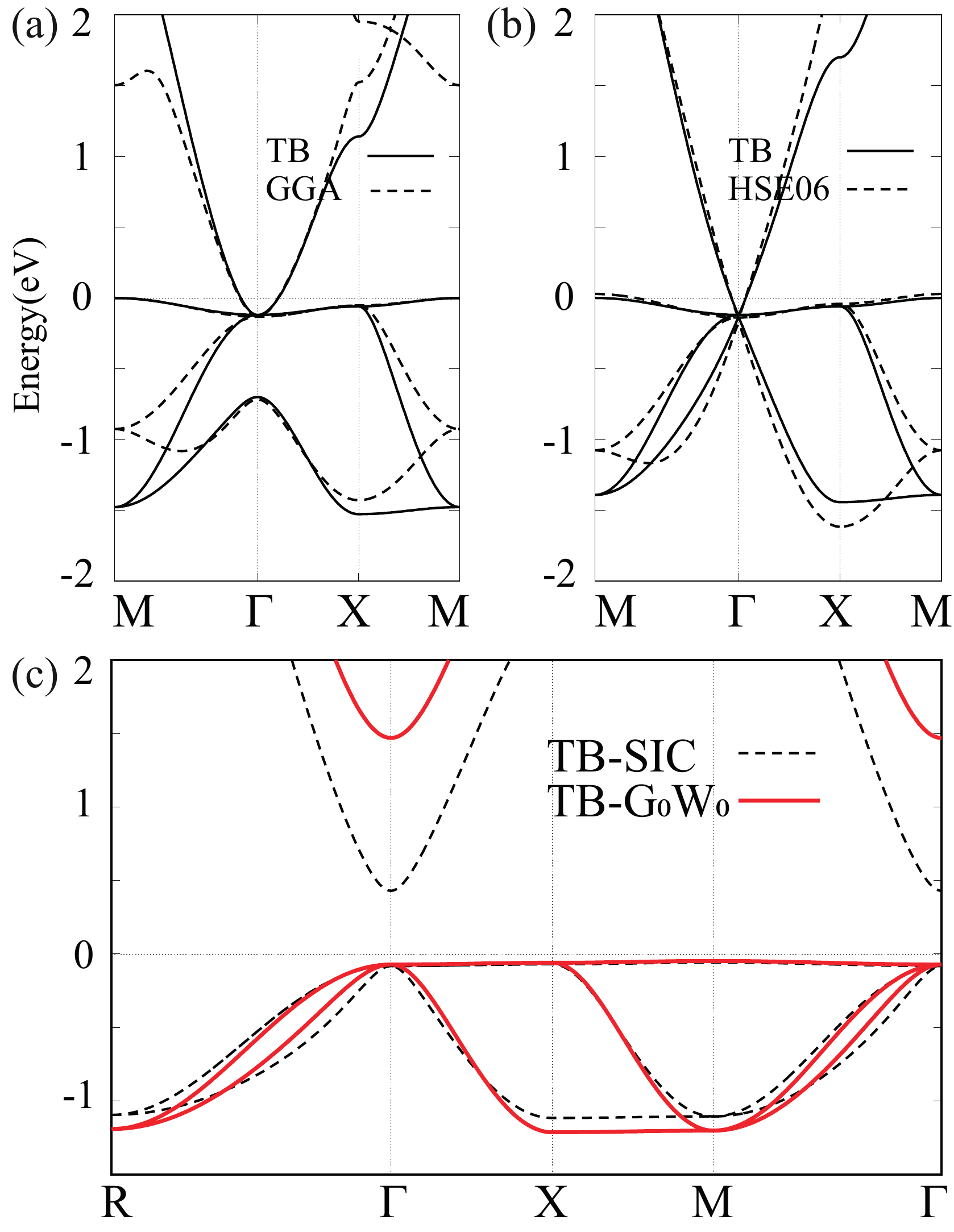} \\
\caption{ (a) GGA band structure (dashed line) of pristine Na$_3$N and
   tight-binding band structure reproducing GGA result (solid line). (b) HSE06
   band structure (dashed line) of pristine Na$_3$N and tight-binding band
   structure reproducing HSE06 result (solid line). (c) Tight-binding band
   structure reproducing the SIC and G$_0$W$_0$ results in \cite{Sommer2012}.}
   \label{fig:sfig1}
\end{figure}

Figures\,\ref{fig:sfig1}(a) and (b) respectively show the band structures of
pristine Na$_3$N obtained by using the GGA and HSE06 exchange-correlation
functionals. Both electronic band structures are calculated as metallic,
similar to the LDA result.  In contrast, literature reports the band structures
of Na$_3$N calculated by using SIC and $G_0W_0$  \cite{Sommer2012}, which are
semiconducting.  These contrasting results reflect a well-known band gap issue
of DFT functionals.  We attribute this disagreement mainly to the different
description for the on-site energies of the Na $s$ orbital.  We find  five sets
of tight-binding parameters that reproduce the each DFT results with the five
different schemes (LDA, GGA, HSE06, SIC, and G$_0$W$_0$) [See
Fig.\,\ref{fig:sfig1} for the GGA, HSE06, SIC and G$_0$W$_0$ band structures].
The on-site energies of the Na $s$ orbitals for the  GGA, HSE06, SIC, and
G$_0$W$_0$ results are increased by 0.00, 0.65, 1.10, and 2.25\,eV, with
respect to the LDA result. The difference of on-site energies $E_{\rm Na,s}$
lead to the semimetallic (semiconducting) band structures of LDA, GGA, HSE06
(SIC, G$_0$W$_0$) $E_{\rm N,p}$. Accordingly, we find a DLN in LDA, GGA, and
HSE06 results under strain, but not in the SIC and G$_0$W$_0$ calculations.  

To resolve the discrepancy of different DFT methods, and to confirm the type-II
DLN semimetal phase in Na$_3$N under strain, we emphasize that a careful set of
new experiments on both pristine and strained Na$_3$N  are crucial. A previous
optical conductivity experiment on the powdered Na$_3$N tentatively reached a
conclusion in favor of an energy gap at the $\Gamma$ point
\cite{VAJENINE2008450}, but we feel that significant refinement in both the
sample preparation and the measurement are still in demand, preferably on
single crystalline sample.

\section*{Appendix B: Effect of SOC}
\begin{figure}[!b]
\centering
\includegraphics[width=0.46\textwidth]{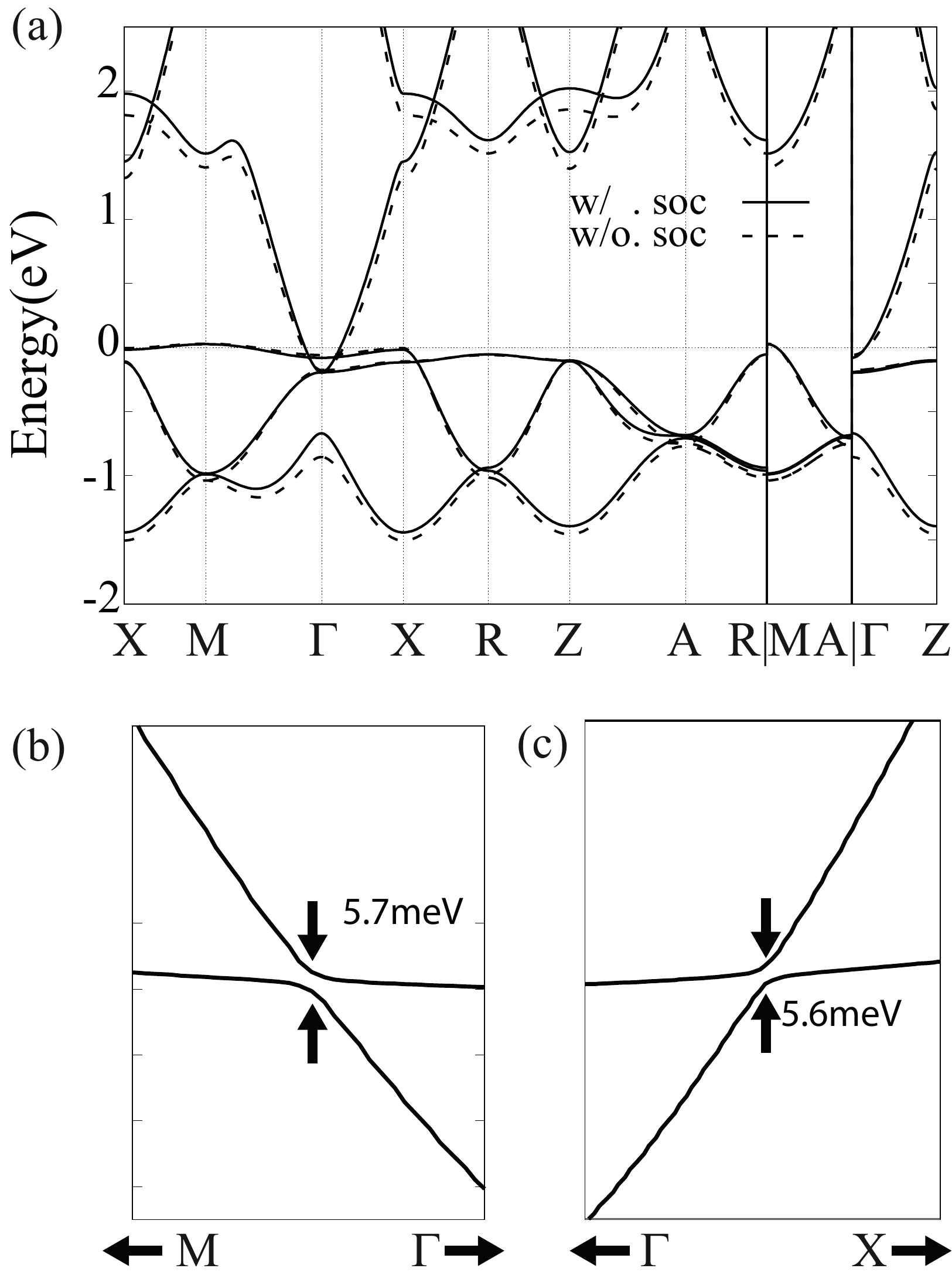} \\
\caption{(a) Electronic band structure of strained Na$_3$N with and with SOC.
   The SOC (non-SOC) band is represented by solid (dashed) line.  The magnified
   views of the band structure on (b) $\Gamma-M$ and (c) $\Gamma-X$. The size
   of the SOC gap on $\Gamma-X$ ($\Gamma-M$) is calculated as $\sim$ 5.6\,meV
   ($\sim$ 5.7\,meV).}
\label{fig:sfig2}
\end{figure}

For the sake of completeness of the study, we calculate the electronic band
structure of Na$_3$N with SOC shown in Fig.\,\ref{fig:sfig2}(a).  Indeed the
effect of SOC is negligibly weak.  It is found that SOC opens a tiny band gap
along the entire DLN.  As shown in Figs.\,\ref{fig:sfig2}(b) and (c), a band
gap opens by $\sim$ 5.6\,meV and $\sim$ 5.7\,meV on the high-symmetry $\Gamma-M$
and $\Gamma-X$ lines, respectively. Considering SOC, strained Na$_3$N can be
considered as a strong topological insulator protected by time-reversal
symmetry. We confirm the nontrivial topological insulator phase induced by SOC
by calculating the $\mathbb{Z}_2$ invariants using parity eigenvalues.  This
calculation results in (1;000), indicating the strong topological insulator.

\section*{Appendix C: Parameters of the $\v {k}\cdot\v {p}$ Hamiltonian}
In Table\,\ref{stable:parameter}, we present the parameters of the $\v k\cdot\v p$ Hamiltonians for 1\,\%,
2\,\%, 3\,\%, and 4\,\%-strained Na$_3$N, which best reproduce the corresponding
first-principles band structures. These parameters were used to calculate the
optical conductivities presented in Sec.\,\ref{subsec:opcd}.
\begin{table*}[tp]
\caption{Parameters of the $\v {k}\cdot\v {p}$ Hamiltonians for 1\,\%, 2\,\%, 3\,\%,
   and 4\,\%-strained Na$_3$N.}
\label{stable:parameter}
\centering
\begin{tabular}{c c c c c}
\hline
Parameters(eV) & 1\,\%-strained & 2\,\%-strained & 3\,\%-strained  & 4\,\%-strained\\
\hline
$a_1$         & 5.00 & 5.00   & 5.00 & 5.00 \\
$b_1$         & 4.60 & 4.60  & 4.60 & 4.60 \\
$a_2$         & 4.82 & 4.64  & 4.46 & 4.28 \\
$b_2$         & 4.46  & 4.32 & 4.18  & 4.04 \\
$v$           & 0.65 & 0.78  & 0.92  & 1.02 \\
$k_D$        & 0.070  & 0.095  & 0.109  & 0.125 \\
\hline
\end{tabular}
\label{table:kp2}
\end{table*}

\end{document}